\documentclass{aa_}
\usepackage{graphicx}
\usepackage{txfonts}
\usepackage{amssymb}
\usepackage{amsfonts}
\usepackage{amsmath}
\usepackage{color}

\newcommand{\fref}[1]{Fig.\ref{#1}}

\newcommand{\eref}[1]{Eq.(\ref{#1})}
\newcommand{\erefs}[1]{Eqs.(\ref{#1})}
\newcommand{\reff}[1]{(\ref{#1})}
\newcommand{\citepp}[1]{(\cite{#1})}

\newcommand{\p}{\partial}
\newcommand{\ie}{\emph{i.e., }}
\newcommand{\no}{\nonumber}

\begin{document}

\title{Coexistence of magneto-rotational and Jeans instabilities in an\\ axisymmetric nebula}

\author{Giovanni Montani\inst{\ref{en},\ref{sap}} \and Giovanni Palermo\inst{\ref{sap}} \and Nakia Carlevaro\inst{\ref{en},\ref{lt}} }

\institute{
     ENEA, Fusion and Nuclear Safety Department, C.R. Frascati, Via E. Fermi 45 (00044) Frascati (RM), Italy \label{en}
\and Department of Physics, ``Sapienza'' University of Rome, P.le Aldo Moro, 5 (00185) Roma, Italy \label{sap}
\and L.T. Calcoli, Via Bergamo, 60 (23807) Merate (LC), Italy \label{lt}}

\date{\phantom{pippo}}

\abstract
{}
{We analyze the magneto-rotational instability (MRI) effects on gravitational collapse and its influence on the instability critical scale.}
{In particular, we study an axisymmetric nonstratified differentially rotating cloud, embedded in a small magnetic field, and we perform a local linear stability analysis, including the self gravity of the system.}
{We demonstrate that the linear evolution of the perturbations is characterized by the emergence of an anisotropy degree of the perturbed mass densities. Starting with spherical growing overdensities, we see that they naturally acquire an anisotropy of order unity in their shape. Despite the linear character of our analysis, we infer that such a seed of anisotropy can rapidly grow in a nonlinear regime, leading to the formation of filament-like structures. However, we show how such an anisotropy is essentially an intrinsic feature of the Jean instability, and how MRI only plays a significant role in fixing the critical scale of the mode spectrum. We then provide a characterization of the present analysis in terms of the cosmological setting, in order to provide an outlook of how the present results could concern the formation of large-scale structures across the Universe.\vspace{0.5cm}}
{}
\maketitle

\section{Introduction}
One of the most intriguing features of magneto-hydrodynamics (MHD) consists of the generation of unstable behaviors from the coupling between wave propagation and plasma inhomogeneities \citepp{Bisk}. A relevant example of such instability mechanisms corresponds to the so-called magneto-rotational instability (MRI), firstly discovered by E.P. Velikhov (\citeyear{Ve59}) and applied to the astrophysical context by S. Chandrasekhar (\citeyear{Ch60}); it results from the coupling of Alfv\'en waves with the differential rotation of the plasma.

It is easy to understand the interest that MRI has raised over the last five decades toward its astrophysical applications, especially since it was clarified \citepp{BH91} (see also \cite{BH98} and \cite{Ba03}) that it constitutes the basic instability able to trigger the turbulence in stellar accretion disks. Indeed, the existence of MRI in accreting structures is crucial to justify the assumption of an effective visco-resistive plasma, which forms part of the basis of the original Shakura idea of accretion onto compact objects \citepp{Sh73,SS73}. Most of the studies on MRI have been locally developed, that is, they consider small wave lengths of the perturbations with respect to the typical variation scale of the background configuration. For a satisfactory discussion of MRI in the global approach, see \cite{Papa92}. The most important results  concerning MRI come from the analysis of the background plasma profiles, according to the idea that vertical shear weakly couples to the perturbation dynamics. Nonetheless, for recent developments concerning MRI in stratified configurations, revising the original approach of \cite{Ba95}, see \cite{MCP16} and \cite{CM17}. Finally, for studies of the morphology of MRI in the presence of dissipative effects, like viscosity and resistivity, we refer to the reviews by \cite{BH98}, and \cite{Sh15} and to \cite{CMR17} for recent investigations.

In this paper, we apply MRI to a different context with respect to stellar accretion disks, to examine whether or not it could play a role in the structure formation across the Universe. Indeed, the plasma nature of the Universe before recombination and the existence of primordial magnetic fields \citepp{Ba07} suggest that the cosmological perturbation dynamics, and therefore the structure formation, can be influenced by MHD effects. In particular, in the recombined-matter-dominated Universe, we can expect that differential rotation significantly enters the nonlinear gravitational collapse of overdensities across the Universe. In this respect, it is important to stress that the Universe remains ionized even after recombination for one part in one hundred thousand. Therefore, the emergence of MRI in the linear and nonlinear phases of the overdensity collapse is a reliable phenomenon for the early Universe, especially due to the tight coupling between ions and neutral hydrogen on many relevant cosmological scales in the era between the recombination and structure formation \citepp{LCM12}. However, the cosmological implementation (provided in Sect. \ref{sec6}) of the present results, obtained for a steady linearized MHD system, has to be regarded as a qualitative hint for further investigations which include the detailed role of the Universe expansion and the relativistic perturbation dynamics.

In \cite{SMC16}, the role played by MRI in the linear stability of a plasma infinite filament is considered in order to investigate possible implications on the morphology of the filament in different regimes. For further studies on the interaction between self-gravitating and magnetic instabilities, see for instance \cite{FBDV04a,FBDV04b,FBDV04c}. Here, we consider the same type of problem, but we refer to a nebula of plasma treated as a differentially rotating nonstratified configuration, in order to evaluate how the corresponding Jeans scale \citepp{jeans} is affected by MRI. The main issue of such an analysis is to demonstrate how, starting with an axisymmetric profile (\ie a configuration of plasma in differential rotation within a weak magnetic field), the linear gravitational collapse is able to induce anisotropic features of the overdensity, up to some percent of the background density and of order unity for the perturbation itself. This linear and local investigation suggests that the extreme nonlinear regime of the collapse could be significantly affected by such initial anisotropies and the emergence of filament-like structures could be justified by the nonlinear evolution of such a linear initial condition. The coupling of the MRI with the Jean mechanism of self-gravitational collapse (\ie the instability due to the presence of differential rotation in a weakly magnetized plasma) could, in principle, influence the process of fragmentation of the nebula since we show that the critical scale of instability is fixed both by gravitational and magneto-rotational effects. However, we clarify how the anisotropy feature of the observed linear growth rates is essentially due to an intrinsic property of the pure Jeans mechanism. Indeed, the largest growth-rate values correspond to small perturbation wave numbers, where MRI is almost suppressed. On the other hand, those wave numbers corresponding to a leading character of MRI exist is the region where the Jeans growth rates strongly decrease: such two instabilities turn out to coexist within the nebula instability profile. For a discussion of similar questions in the general relativistic sector, see \cite{GT06}, where the evolution of small linear anisotropies is investigated in a specific case, demonstrating their enhancement by the gravitational collapse.

\section{Basic equations}
Let us consider a self gravitating plasma axial structure (nebula) described in standard cylindrical coordinates $(r,\,\varphi,\,z)$, whose self-gravity potential is denoted by $\phi$. The nebula is taken in differential rotation with angular velocity $\Omega=\Omega(r)$, and it is embedded into a purely vertical magnetic field $\boldsymbol{B}_0=B_0\hat{\boldsymbol{e}}_z$. Since we are considering a weak dependence on the vertical direction of all the background quantities, they are taken as a function of $r$ only and the nebula stratification is, on this level, neglected.

The theory properly describing the physics governing such a system is therefore the ideal MHD, whose fundamental equations are the mass, momentum, and magnetic flux conservation laws
\begin{align}
\p_t\rho+\nabla\cdot(\rho\boldsymbol{v})=0\;,\label{eq:62}\\
\rho\p_t\boldsymbol{v}+\rho(\boldsymbol{v}\cdot\nabla)\boldsymbol{v}+\nabla(p+B^{2}/8\pi)\qquad\qquad\quad\no\\
-\rho\nabla\phi-(\boldsymbol{B}\cdot\nabla)\boldsymbol{B}/4\pi=0\;,\\
\p_t\boldsymbol{B}-\nabla\wedge(\boldsymbol{v}\wedge\boldsymbol{B})=0\;,
\end{align}
respectively. Here, $\rho$ denotes the mass density of the fluid, $p$ the hydrostatic pressure, while $\boldsymbol{v}$ and $\boldsymbol{B}$ ($B=|\boldsymbol{B}|$) are the velocity and magnetic field, respectively. In addition, we consider the Poisson equation describing the self-gravity of the nebula, that is,
\begin{align}
\nabla^{2}\phi-4\pi G\rho=0\;, 
\end{align}
with $G$ being the Newton constant and, finally, we specify the equation of state (EoS) as
\begin{align}\label{eq:eos}
p-v_{s}^{2}\rho=0\;,
\end{align}
where $v_s$ is the sound speed and we consider an isothermal relation between pressure and mass density ($v_s\simeq const.$).

\subsection{Background equilibrium}
Let us now describe the fundamental equations governing the time-independent background equilibrium, characterized by a set of variables indicated via the subscript $0$, in order to distinguish from the dynamical perturbed quantities of the following section denoted by a subscript 1.

Assuming that the background magnetic field $\boldsymbol{B}_0$ is sufficiently small to have no effect on the steady equilibrium of the plasma, we can easily fix the equation governing its gravostatic configuration as
\begin{align}
r\Omega^{2}(r)-(\p_r p_0)/\rho_{0}-\p_r\phi_0=0\;,\\
(\p_r(r\p_r\phi_{0}))/r-4\pi G\rho_{0}(r)=0\;,\\
\p_r p_0-v_s^2\p_r\rho_0=0\;.\label{eq:6}
\end{align}
Here, we neglected the $z$ dependence of the problem, while all the $\varphi$ derivatives vanish because of axial symmetry. Actually, the $z$ dependence is assumed negligible here, bearing in mind that the perturbations will have a small wavelength, making it difficult to explore the vertical shear of the configuration. Nonetheless, the role of the vertical gradients can be, in principle, important in order to fix the background profile. In this case, the system above must include vertical gradients as well as the vertical force balance.

\section{Linear perturbation theory}
Let us now face the problem of characterizing the linear stability of the considered plasma configuration. In what follows, we consider a local approximation, that is, the typical perturbation length is considered much smaller than the length scale of the background variation. The background quantities and their radial derivatives are taken at a fiducial radius $r=\bar{r}$, and they are treated as constants in the considered perturbative problem. According to this approximation, the perturbation equations, associated to the system of \erefs{eq:62}-\reff{eq:eos}, are written as 
\begin{align}
\p_t\rho_{1}+\rho_{0}(\nabla\cdot\boldsymbol{v}_{1})+\boldsymbol{v}_0\cdot\nabla\rho_{1}=0\;,\label{qe:ieoe1}\\
\rho_{0}(\p_t\boldsymbol{v}_1+(\boldsymbol{v}_0\cdot\nabla)\boldsymbol{v}_{1})+
\nabla(p_{1}+2\boldsymbol{B}_{0}\cdot\boldsymbol{B}_{1}/8\pi)\quad\no\\
-\rho_{0}\nabla\phi_{1}-(\boldsymbol{B}_{0}\cdot\nabla)\boldsymbol{B}_{1}/4\pi=0\;,\\
\p_t\boldsymbol{B}_{1}+\boldsymbol{B}_{0}(\nabla\cdot\boldsymbol{v}_{1})-
(\boldsymbol{B}_{0}\cdot\nabla)\boldsymbol{v}_{1}\qquad\qquad\no\\
+(\boldsymbol{v}_{0}\cdot\nabla)\boldsymbol{B}_{1}=0\;,\\
\nabla^{2}\phi_{1}-4\pi G\rho_{1}=0\;,\\
p_{1}-v_{s}^{2}\rho_{1}=0\;,\label{qe:ieoe2}
\end{align}
respectively.

Since the background does not depend on time, we search WKB solutions writing each axisymmetric perturbation quantity $A_1$ as $A_{1}(t,r,z)=\bar{a} e^{i (k_r r+k_z z-\omega t)}$ with $\bar{a}=const.$ and the wave vector as $\boldsymbol{k}=(k_r,\,0,\,k_z)$. In this way, the following natural replacements take place: $\nabla A_1= i\boldsymbol{k} A_1$ and $\p_t A_1=-i\omega A_1$. The system of \erefs{qe:ieoe1}-\reff{qe:ieoe2} can be rewritten as
\begin{align}
-i\omega\rho_{1}/\rho_{0}+ik_{r}v_{1r}+ik_{z}v_{1z}=0\;,\\
-i\omega v_{1r}-2\Omega v_{1\phi}+ik_{r}p_{1}/\rho_{0}+ik_{r}\phi_{1}\qquad\quad\quad\no\\
+ik_{r}v_{A}^{2}B_{1z}/B_{0}-ik_{z}v_{A}^{2}B_{1r}/B_{0}=0\;,\\
\kappa^{2}v_{1r}/2\Omega-i\omega v_{1\phi}-ik_{z}v_{A}^{2}B_{1\phi}/B_{0}=0\;,\\
-i\omega v_{1z}+ik_{z}p_{1}/\rho_{0}+ik_{z}\phi_{1}=0\;,\\
-i\omega B_{1r}-ik_{z}B_{0}v_{1r}=0\;,\\
-i\omega B_{1\phi}-B_{1r}\p_{\ln r}\Omega-ik_{z}B_{0}v_{1\phi}=0\;,\\
-i\omega B_{1z}+ik_{r}B_{0}v_{1r}=0\;,\\
-k^{2}\phi_{1}-4\pi G\rho_{1}=0\;,\\
p_{1}-v_{s}^{2}\rho_{1}=0\;,
\end{align}
where we introduced the standard definitions of the epicyclic frequency and Alf\'en
velocity
\begin{align}
\kappa^{2}=4\Omega^{2}+\p_{\ln r}\Omega^{2}\;,\qquad
v_{A}^{2}=B_{0}^{2}/4\pi\rho_0\;,
\end{align}
respectively.

It is worth reiterating that the Rayleigh criterion \citepp{BH98}, in the absence of differential rotation, states that the nebula is unstable if $\kappa^{2}<0$. Therefore, in order to select the role played by MRI in the stability of the system, in what follows we take positive epicyclic frequencies $\kappa^{2}>0$. The system above is linear algebraic and homogeneous in the perturbed quantities, and therefore it admits a nontrivial solution only if the corresponding determinant vanishes, therefore leading to the dispersion relation
\begin{multline}
\omega^{6}-\omega^{4}(\kappa^{2}+\omega_{0}^{2}+\omega_{A}^{2}+\omega_{Az}^{2})+\\	+\omega^{2}\big(\omega_{Az}^{2}(2\omega_{0}^{2}+\omega_{A}^{2}+\p_{\ln r}\Omega^{2})+
\kappa^{2}\omega_{0}^{2}\omega_{Az}^{2}/\omega_{A}^{2}\big)\\
-\omega_{0}^{2}\omega_{Az}^{2}\big(\omega_{Az}^{2}+(\p_{\ln r}\Omega^{2})\omega_{Az}^{2}/\omega_{A}^{2}\big)=0\;.\label{eq:236}
\end{multline}
Here, to simplify the notation, we introduce the following frequencies: $\omega_{0}^{2}=v_{s}^{2}k^{2}-4\pi G\rho_{0}$ is the typical frequency appearing in the Jeans (self-gravitation) instability, $\omega_{A}^{2}=v_{A}^{2}k^{2}$ is the Alfv\'en frequency and $\omega_{Az}^{2}=v_{A}^{2}k_{z}^{2}$ represents the Alfv\'en parameter $(\boldsymbol{k}\cdot(v_A \hat{\boldsymbol{e}}_z))^{2}$, which modulates the magnetic tension.

We can now rewrite \eref{eq:236} as a function of the angle $\theta$ between $\boldsymbol{k}$ and $\boldsymbol{B}_0$. Therefore, we define $\chi=\omega_{Az}^{2}/\omega_{A}^{2}=k_{z}^{2}/k^{2}=\cos^{2}\theta$ and, substituting this expression in \eref{eq:236}, we finally get
\begin{multline}
\omega^{6}-\omega^{4}(\kappa^{2}+\omega_{0}^{2}+\omega_{A}^{2}(1+\chi))+\\
+\omega^{2}\big(\omega_{A}^{2}(2\omega_{0}^{2}+\omega_{A}^{2}+\p_{\ln r}\Omega^{2})+\kappa^{2}\omega_{0}^{2}\big)\chi\\
\qquad\qquad-\omega_{0}^{2}\omega_{A}^{2}\big(\omega_{A}^{2}+\p_{\ln r}\Omega^{2}\big)\chi^{2}=0\;,\label{eq:237}
\end{multline}
where, clearly, $0\leqslant \chi \leqslant 1$. We stress that, turning off gravity, \ie $G\to0$, taking $v_{s}^{2}\to\infty$ (which is the limit toward the Boussinesq approximation \citepp{BH98}) and fixing $\boldsymbol{k}=k_z \hat{\boldsymbol{e}}_z$, the expression above reduces to the dispersion formula for the MRI \citepp{BH91}.

\section{Discussion of the dispersion relation}
Before analyzing the solutions of the obtained dispersion relation in detail, we investigate some relevant simplified cases.

\subsection{The limit $\chi=0$}
We start by considering a perturbation $\boldsymbol{k}=k_r\hat{\boldsymbol{e}}_r$, implying $\chi=0$. The dispersion relation takes the following simplified form.
\begin{equation}
\omega^{2}-(\kappa^{2}+\omega_{0}^{2}+\omega_{A}^{2})=0\;.
\end{equation}
The instability condition $\omega^{2}<0$ gives
\begin{equation}
k_r^{2}<\bar{k}_J^{2}\equiv\frac{4\pi G\rho_{0}-\kappa^{2}}{v_{s}^{2}+v_{A}^{2}}\label{eq:239}\;,
\end{equation}
for which the associated Jeans length is
\begin{equation}
\bar{\lambda}_{J}\equiv\frac{2\pi}{\bar{k}_J}=2\pi\sqrt{\frac{v_{s}^{2}+v_{A}^{2}}{4\pi G\rho_{0}-\kappa^{2}}}\label{eq:240}\;.
\end{equation}
This solution is physically meaningful only if $\kappa^{2}<4\pi G\rho_{0}$: when this relation is satisfied, \eref{eq:240} provides the Jeans scale for a rotating, self-gravitating, and magnetized plasma in correspondence to a radial propagation of the perturbations (absence of magnetic tension). Furthermore, the limit $\kappa^{2}\to0$ and $v_{A}^{2}\to0$ reproduces the standard Jeans criterium.

From \eref{eq:240}, we see that the magnetic field contrasts the collapse, since the corresponding Jeans length is larger than the nonmagnetized one. The epicyclic frequency of the fluid contrasts the collapse too and can make the system stable for any perturbation if $\kappa^{2}\geqslant 4\pi G\rho_{0}$. This squared frequency is always intended to be positive, in order to satisfy the Rayleigh criterion. However, negative values for $\kappa^{2}$ can increase the instability of the system, inducing a collapse at smaller wavelengths.

\subsection{Perturbative solution of the dispersion relation}
Since the aim of our analysis is to verify the role played by MRI in the stability of a self-gravitating medium, we consider \eref{eq:237} nearby a static gas cloud by turning on a small magnetic field and epicyclic frequency. More precisely, this means $\omega_{0}^{2}\gg\omega_{A}^{2}$ and $\omega_{0}^{2}\gg\kappa^{2}$. Moreover, we take $\chi\ll1$ in order to have small magnetic tension. The differential rotation parameter is taken at half magnitude between $\omega_{0}^{2}$ and the other frequencies to preserve its relevant role in the MRI instability, \ie $\omega_{0}^{2}>\p_{\ln r}\Omega^{2}\gg\omega_{A}^{2},\,\kappa^{2}$.

Under these assumptions, keeping only the first-order perturbations, the dispersion formula can be rewritten as
\begin{equation}
\omega^{4}-\omega^{2}\left[\kappa^{2}+\omega_{0}^{2}+\omega_{A}^{2}\left(1+\chi\right)\right]+\frac{\omega_{A}^{2}}{\omega_{0}^{2}}\,\p_{\ln r}\Omega^{2}=0\;.
\end{equation}
Considering the solution with the $+$ sign, because it represents a modified Jeans frequency, we get
\begin{equation}
\omega^{2}=\omega_{0}^{2}+\kappa^{2}+\omega_{A}^{2}\left(1+\chi\right)+
\frac{\omega_{A}^{2}}{\omega_{0}^{2}}\,\p_{\ln r}\Omega^{2}\;.
\end{equation}
Given $\omega_{0}^{2}<0$ , the role of MRI naturally emerges. As magnetic tension and differential rotation are turned on, the root tends to be less negative if $\p_{\ln r}\Omega^{2}<0$, which is the condition for MRI. Therefore, the obtained instability mitigates the pure Jeans one. Instead, a positive differential rotation parameter makes the system even more unstable.

\subsection{General case}
Let us now analyze the dispersion relation by plotting its numerical solutions in terms of the model parameters. In what follows, we expect to be able to recognize, in the profile of the growth rate $\gamma$ (the positive imaginary part of $\omega$) as a function of $k$, both the MRI and the Jeans behavior. Naming the unstable solutions $\omega_{J}^{2}$ and $\omega_{M}^{2}$ (since their behaviors match Jeans and MRI modes, respectively), we consider the hybrid root defined as $\omega_{M\!J}^{2}(k)=\textrm{min}[\omega_{J}^{2},\,\omega_{M}^{2}]$, where $MJ$ stands for magneto-Jeans. The critical wavelength of the system is therefore set as $\lambda_{M\!J}=\textrm{min}[\lambda_{J},\,\lambda_{M}]$.
\begin{figure}[ht]\centering
\includegraphics[width=0.8\columnwidth]{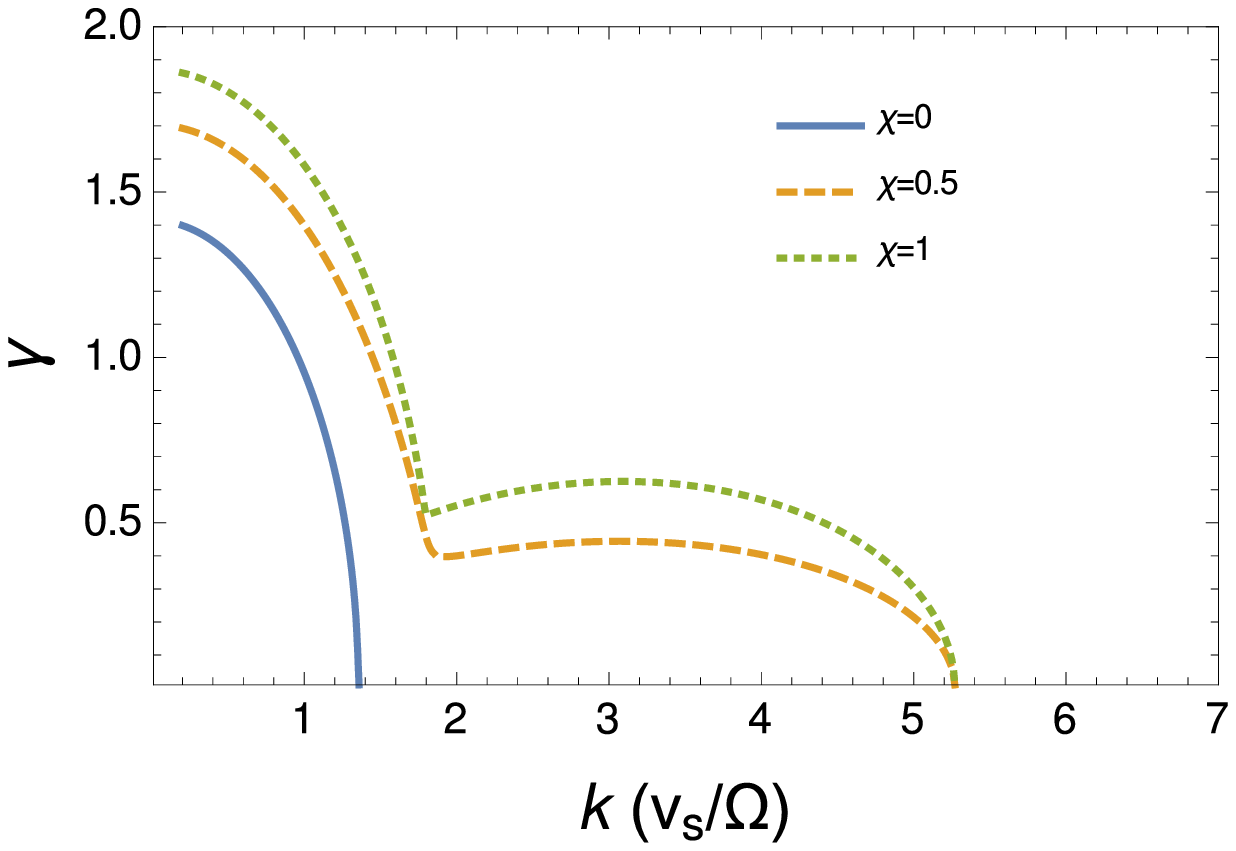}\\
\includegraphics[width=0.8\columnwidth]{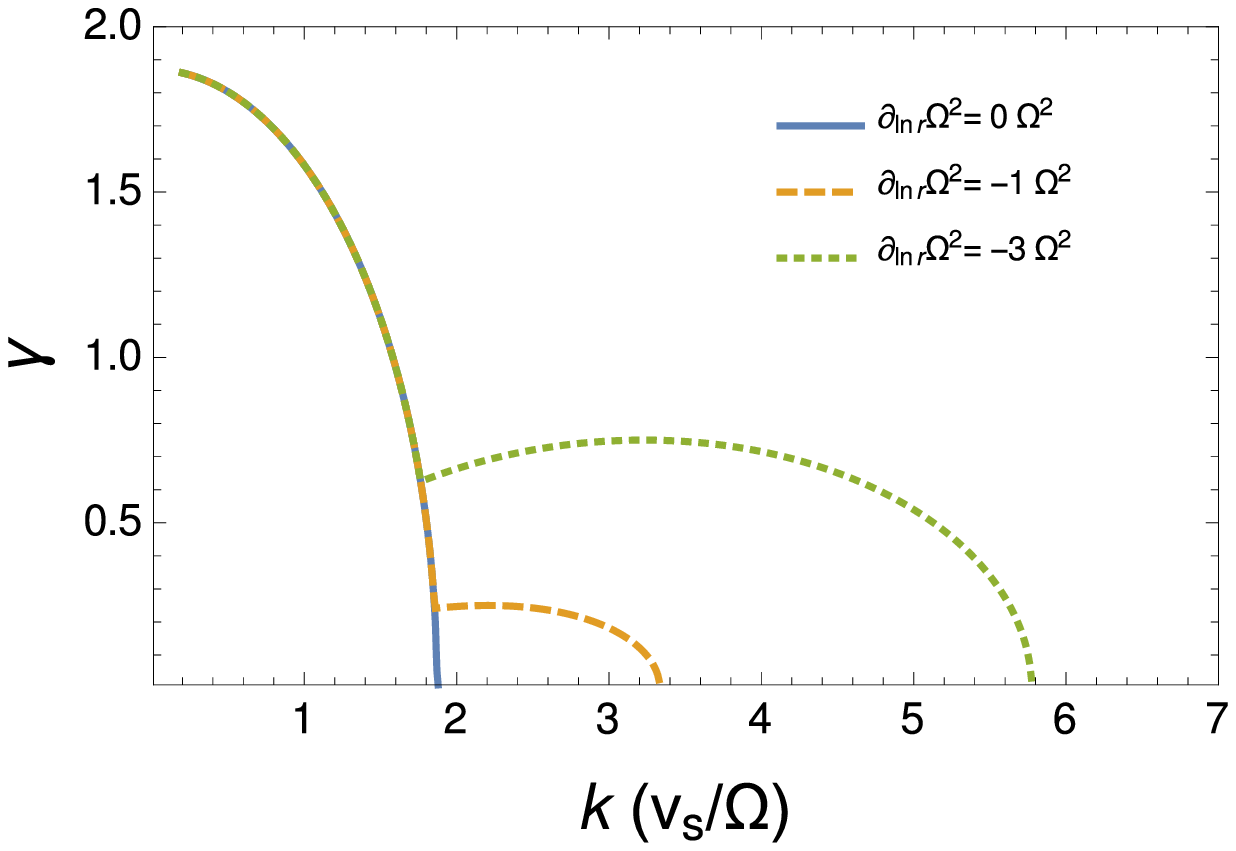}
\caption{MJ growth rate (in $\Omega$ units) from \eref{eq:237} as a function of $k (v_s/\Omega)$ for different values of the wave vector angle fixing $\p_{\ln r}\Omega^{2}=-2.5\Omega^2$ (upper panel), and for different values of the differential rotation fixing $\chi=1$ (lower panel), as indicated in the plot. Integration parameters are: $v_A/v_s=0.3$ and $4\pi G\rho_0=3.5\Omega^2$. For $\chi=0,$ the curve reduces to the one associated to the standard Jeans instability, while for $\chi\neq0,$ a second peak arises. This effect is increased by differential rotation. The critical wavelength is discontinuous for $\chi=0$ and $\chi\neq0$ (where it remains constant changing the value of $\chi$).\label{fig:1}}
\end{figure}

As can be inferred by the upper panel of \fref{fig:1}, the collapse (the growth of the mass density) is significantly anisotropic since it depends on the ratio $\chi=k_z^2/k^2$. Anisotropy is also increased by differential rotation, as can be argued from the lower panel of the figure, which, moreover, pushes the critical wavelength to smaller scales (however, we note that $\p_{\ln r}\Omega^2$ must always be negative in order to have an MRI contribution \citepp{BH91}). Moreover, as can be seen from the upper panel of \fref{fig:1}, no dependence of instability critical scale on the value of $\chi$ emerges, except the discontinuity between $\chi=0$ and $\chi\neq0$.

\begin{figure}[ht]\centering
\includegraphics[width=0.8\columnwidth]{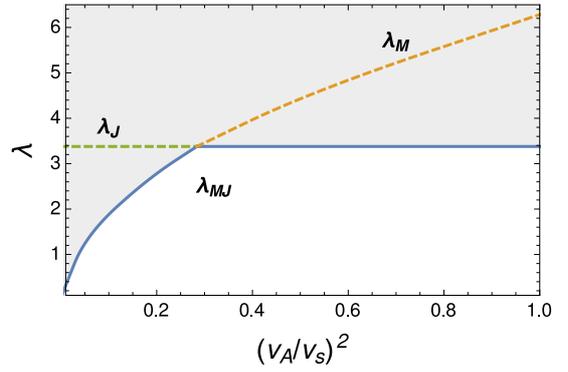}
\caption{MJ critical length $\lambda_{M\!J}=\textrm{min}[\lambda_{J},\,\lambda_{M}]$ (as indicated in the plot) as a  function of the ratio $(v_A/v_s)^2$. The parameters are set as: $4\pi G\rho_0=3.5\Omega^2$, $\chi=0.5$ and $\p_{\ln r}\Omega^{2}=-1\Omega^2$. \label{fig:2}}
\end{figure}

In \fref{fig:2}, the critical length $\lambda_{M\!J}$ is plotted as a function of the ratio $(v_A/v_s)^2$, associated to the change of the magnetic field strength amplitude.
We note that MRI is valid for weak magnetic field only, but it implies no matter transport, and, if coupled with self-gravity, it extends the gravitational collapse to smaller scales, as clarified in the figure.

It is worth noting that the upper panel of \fref{fig:1} shows that, for a given $k$, the growth rate is larger for modes with $\chi=1$ (dotted line) than for modes with $\chi=0$ (solid line). Moreover, as clearly deduced from  \fref{fig:2}, at small values of $k$ (large scales), the dominant instability is the Jeans one, whereas for large $k$ (small scales), the dominant instability is associated to the MRI contribution. Below, we see how these considerations concern the analysis of the density-contrast evolution.

\section{Time evolution of the density contrast}
Let us now estimate the amount of the collapse anisotropy by studying the behavior of the density contrast defined as $\delta(\boldsymbol{r},t)=\rho_{1}/\rho_{0}$ (where $\boldsymbol{r}=(r\hat{\boldsymbol{e}}_r+z\hat{\boldsymbol{e}}_z)$ is the poloidal vector radius). We consider an initial overdense region of Gaussian form, \ie
\begin{equation}
\delta(\boldsymbol{r},0)=\delta_{0}e^{-\frac{r^{2}+z^{2}}{2\sigma}}\;,
\end{equation}
whose Fourier transform is still Gaussian, 
\begin{equation}
\tilde{\delta}(\boldsymbol{k},0)=\tilde{\delta}_{0} e^{-\frac{k_{r}^{2}+k_{z}^{2}}{2\tilde{\sigma}}}\;;
\end{equation}
above, $\sigma$ and $\tilde{\sigma}$ are the direct space and Fourier space variance, being reciprocal numbers ($\delta_{0}$ and $\tilde{\delta}_{0}$ are assigned constants).

In the $\boldsymbol{k}$-space, the single mode evolution is described by
\begin{align}\label{wsjnsjun}
\tilde{\delta}(\boldsymbol{k},t)=\tilde{\delta}_{0}
e^{-\frac{k^{2}}{2\tilde{\sigma}}+i\boldsymbol{k}\cdot\boldsymbol{r}-i\omega_{M\!J}\,t}\;,
\end{align}
where $\omega_{M\!J}=\omega_{M\!J}(\boldsymbol{k})$ is the MJ frequency derived in the previous section as a solution of the dispersion relation. Taking $\lambda>\lambda_{M\!J}$,  $\omega_{M\!J}=i\gamma_{M\!J}$ and perturbations grow exponentially. Transforming back \eref{wsjnsjun} into the $\boldsymbol{r}$-space, we can obtain the density contrast as a function of position until $\delta\ll1$. In fact, when $\rho_1\simeq\rho_0$, the linear analysis is clearly no longer valid and simulations of the nonlinear problem are required.

\subsection{Evolution of the overdensity}
\begin{figure}[ht]\centering
\includegraphics[width=0.7\columnwidth]{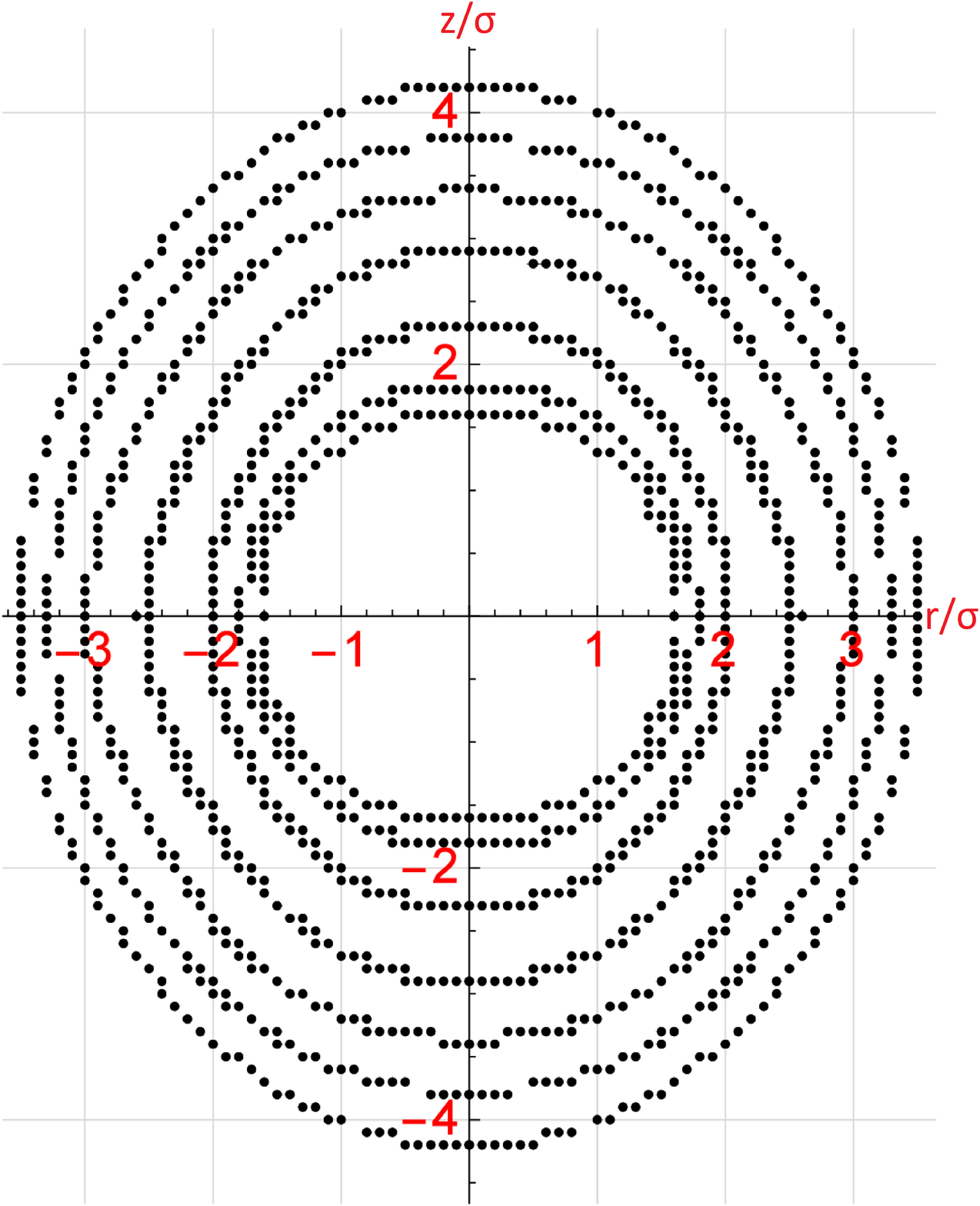}
\caption{\label{fig:3.23} Discrete representation in the plane $(r,\,z)$ of the isolines $\delta=0.1$ at $t\,\Omega=0$ (inner region), $0.25,\,0.5,\,0.75,\,1,\,1.25,\,1.5$ (outer region).}
\end{figure}

The wave packet describing the evolution of the  overdense region is now evaluated for different values of $\boldsymbol{r}$. In this way, it is possible to investigate the dynamics of the collapse geometry by plotting the isolines of $\delta(\boldsymbol{r},t)$ on the $r$-$z$ plane at different times (the system is invariant under translation along $\varphi$). The anisotropic behavior of the overdensity growth rate is evident from its dependence on the angle $\chi,$ and therefore we see the distortion of the isolines as time goes by, according to  \fref{fig:3.23}. Here, the numerical integration of the packet is performed taking $\tilde{\delta}_{0}=0.1\times e^{-k^2/2}$.

When $t=0$, the isoline corresponds to a circle, but the shape is progressively altered as time increases, because the growth rate is larger along the direction of the background magnetic field. In fact, in Fig.\ref{fig:3.23}, the isoline $\delta=0.1$ stretches along the $z$ axis. Given a distance $d$, $\delta(r=d,z=0,t)<\delta(r=0,z=d,t)$, \ie the blob is more dense on the $z$ axis and therefore it prevalently compresses in that direction. Furthermore, we note that the discontinuity in the growth rate derivative (see Fig.\ref{fig:1}) has no visible effects on the density contrast, although this bend corresponds to a jump in the group velocity $\p_k \omega_{M\!J}$.

The shape of the isolines does not change if we choose another value for $\tilde{\delta}_{0}$, since it simply changes a constant that multiplies the Fourier transformation. We observe that we cannot overcome $\tilde{\delta}_{0}\sim0.1$, otherwise we violate $\rho_{1}\ll\rho_{0}$ and the linear approximation is no longer valid.

\subsection{Evaluation of the anisotropic growth}
Let us now construct a coefficient in order to evaluate the anisotropy of the growing instability:
\begin{equation}
\alpha(t)=\frac{\delta(r=0,\,z=d,\,t)}{\delta(d,\,0,\,t)}\;,
\end{equation}
where $d$ is a value of the distance from the center fixed with respect to the isoline $\delta=0.1$ on the $z$ axis and at a given time, namely $\delta(0,\,d,\,1.25)=0.1$. We can see that the quantity $\alpha$ can reach large values; for instance, at the chosen instant $t=1.25$, we get $\alpha(1.25)\simeq0.5$. It must be stressed that $\alpha$ measures the anisotropy of the density contrast only, but it allows us to evaluate the whole density anisotropy. In fact, we can define an anisotropy coefficient for the whole density as
\begin{equation}
A(t)=1-\frac{\rho(d,\,0,\,t)}{\rho(0,\,d,\,t)}\;,
\end{equation}
and therefore this coefficient reaches $1$ (and the anisotropy is of the order $100\%$) for $\rho(d,\,0,\,t)/\rho(0,\,d,\,t)\to0$; it can be rewritten as function of $\alpha$ as
\begin{equation}
A(t)=1-\frac{1+\alpha\delta(0,\,d,\,t)}{1+\delta(0,\,d,\,t)}\;.
\end{equation}
It is easy to evaluate that $A(1.25)\simeq5\%$, meaning that the onset collapse is just slightly anisotropic according to the considered linear regime ($\rho_1\ll\rho_0$).

It important to stress here that if we consider an initial perturbation where all modes are excited, the modes that grow faster are those with small $k$ values, since the corresponding growth rate is larger (see \fref{fig:1}). Here the MRI is essentially irrelevant and therefore the anisotropy feature, \ie the difference between the dashed line and dotted line in \fref{fig:1}, is only due to the fact that the Jeans mechanism exhibits different growth rates for modes with different $\chi$ values. We therefore deduce that magnetic tension is unimportant for sufficiently large scales and we are obtaining an equivalent behavior of the overdensity to the one predicted by \eref{eq:237} in the limit $\omega_A\to0$ and $k\to0$. On the other hand, if we focus our attention on small scales, for which the Jeans instability is suppressed, then only the MRI is relevant, and, in this case, the growth rate is again anisotropic (clearly the MRI is strongly suppressed for $\chi\to0$). Nonetheless, this effect is essentially uncoupled from the Jeans instability and the gravitation and magneto-rotational regions of the unstable spectrum simply coexist within the differentially rotating nebula profile.

\section{Phenomenological considerations}\label{sec6}
Our investigation on the co-existence of MRI and Jeans instability and the main conclusion we reach about the noninteraction of the two mechanisms, which however offer a scenario for the generation of anisotropic structures, can be applied in different physical contexts, corresponding to different scales and systems across the Universe. The most natural implementation of the present study is in those astrophysical systems, such as filaments in the interstellar medium and nebula-like structures, which possess a sufficiently high level of ionization and self-gravity to be interpreted via the instability features, predicted by the dispersion relation \reff{eq:237}.

Nonetheless, we aim to infer the validity of the present study on a cosmological level, in order to characterize the behavior of small Universe inhomogeneities in the evolution range between the hydrogen recombination ($z\simeq 1100$) and the formation of the most common structures at large scales ($z\sim 10$), where $z$ denotes the \emph{red-shift}. However, in order to successfully address this characterization of  \eref{eq:237}, we must consider some subtle questions concerning the primordial Universe: 1) the Universe background is expanding, \ie it is nonstationary, and it is also homogeneous in space; 2) only a very weak part of the Universe baryonic component is ionized, about a one part in one hundred thousand; 3) the Universe
also contains dark energy (about $70\%$ of its total energy density) and dark matter (about $25\%$ of its energy density).
 
We now consider each of these questions separately point by point, arguing how the present study is \emph{de facto} applicable to the early Universe, providing qualitative but reliable information about the dynamics of its inhomogeneities:

\begin{enumerate}
\item The nonstationarity of the Universe is relevant for the physics of the early cosmology, including the behavior of linear perturbations, only for spatial scales comparable with the Hubble size ($L_H\simeq c H^{-1}(t)$, where $H(t)$ is the expansion rate and $c$ the speed of light). Such a scale roughly increases as $L_H\sim ct$,  $t$ being the universal synchronous time, and it is well-known \citepp{KT90,M2011} that many cosmological scales, relevant for structure formation, become smaller than $L_H$ simply because they increase slower in time, according to the cosmic scale factor behavior $a(t)\sim t^{2/3}$. Therefore, when studying cosmological perturbations whose size is well below the Hubble length $L_H$, the effects of the expansion can be safely neglected and the background can be regarded as a steady one. 

An important feature introduced by the expansion, with respect to the steady case here considered, consists of a power-law growth in time of the perturbation, against an exponential instability. However, as well-known in the nonmagnetized case \citepp{M2011,Wein72} and also validated in the presence of a magnetic field in \cite{LCM12} and \cite{PD12} (see also \cite{VTP05}), the concept of Jeans threshold scale can be defined in both cases (without or with expansion, respectively). Furthermore, it comes out that the value of the Jeans scale in the two cases differs for a numerical factor only. The presence of a non-stationary expansion therefore does not affect the physical content of our linear analysis, but it could affect the timescale of the considered processes.

However, our request of a cylindrical symmetry, implied by the presence of differential rotation (we assume the existence of a privileged direction), suggests that our analysis must essentially concern the stability of subregions inside primordial structures, \ie the background must be considered to be density contrast close to the nonlinear regime, almost bounded and disconnected by the Hubble flux. In this respect, our study can be regarded as relevant for the iteration of the so-called Jeans mechanism for the Universe fragmentation, especially around $z\sim100$, where the density contrast can be greater than unity and the dark matter role is not relevant to the jet (see below point 3).

\item Although the Universe is weakly ionized after the recombination (actually the Universe expansion prevents perfect hydrogen recombination), we can nonetheless reliably argue that the behavior of the ionized component is closely linked to the neutral matter behavior, meaning that our analysis can be applied to the baryonic neutral Universe as well, for a wide range of cosmological scales.

Ions and neutral atoms actually interact via collisions, mediated by a drag coefficient $\gamma_{in}\sim 1.9\times 10^{-9} $cm$^3$s$^{-1}$. The relevance of this interaction process, commonly dubbed \emph{ambipolar diffusion} (see, e.g., \cite{SS05,SS08,Sch09a,Sch09b}), is properly characterized by the ambipolar Reynold number $R_{amb}$, defined as follows \citepp{LCM12}
\begin{equation}
R_{amb} \equiv L/L_{amb}\;,\qquad
L_{amb} \equiv v_A^2/v\gamma_{in}n_i\;, 
\label{amb1}
\end{equation}
where $n_i$ is the ion number density and $v$ and $L$ are a typical velocity and spatial scale of the considered system, respectively. In the present study, $v\sim v_s\gtrsim v_A$, since the MRI typically holds for high values of the plasma $\beta$ parameter.

It is well-known \citepp{MS56,Sh83,BJ04,Li06}, that when $R_{amb}\ll 1,$ the ions and neutral atoms are very weakly interacting and we deal with two distinct components in the system; when $R_{amb}\sim 1$, the two components are coupled and the ambipolar diffusion is relevant, being represented by a dissipative term in MHD (the single fluid representation holds, but a correction to ideal case must be taken into account); finally, for $R_{amb}\gg1$, that is,  $L\gg L_{amb}$, the coupling becomes very strong and the two species are so tightly evolving that the ideal MHD representation can be applied for the whole system, in agreement with the present study.

It is possible to show (see Fig.1 in \cite{LCM12}) that at the Universe background density for $10<z<1100$, the Reynold number $R_{amb}$ remains much larger than unity for spatial scales containing a mass much greater than $10^6$ solar masses, \ie for the most relevant cosmological scales. Since we intend to apply our analysis to the stability of overdense regions across the Universe, the situation is even slightly better, because $L_{amb}$ is correspondingly smaller. Therefore, we can safely assume to be in a parameter region where the ambipolar diffusion dissipation term is actually negligible, but ions and neutrals are tightly bound forming a single fluid.

In order to characterize the cosmological scales for which our analysis is predictive, in \fref{fig:rcc} we plot, as function of the red-shift $(1+z),$
\begin{figure}[ht]\centering
\includegraphics[width=0.7\columnwidth]{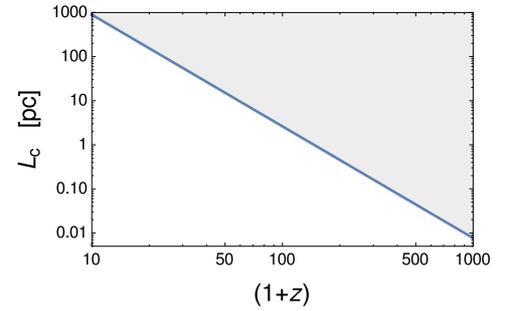}
\caption{\label{fig:rcc} Critical scale $L_c \Rightarrow R_{amb}=1$ as function of the redshift for $4\pi G\rho_0=3.5\Omega^2$. Marked region denotes scales implying $R_{amb}>1$.}
\end{figure}
the size associated to the minimal mass of the structure to deal with $R_{amb}>1$ (we define the scale $L_c$ which imply $R_{amb}=1$), which corresponds to the value $3.5$ of the ratio $y_g\equiv4\pi G\rho_0/\Omega^2$, used in the plots of this paper and reasonable for a cylindrical configuration (cf. with Fig.1 in \cite{LCM12}). The field intensity $B_{0}(z)$ behaves as $(1+z)^{2}$ and we set its present value as $B_0(z=0)=10^{-9}$G. Since we are considering our background to be an overdensity of the expanding Universe, we consider $\rho_0=3\rho_{crit}(1+z)^3$ ($\rho_{crit}$ being the Universe critical density): the factor $3$ causes the considered region to be almost disconnect from the Hubble flux. We remark that $L_c$ is defined by $L_c^3=M_c y_g/4\pi\rho_0$, where $M_c(z)$ denotes the critical mass contained within the scale $L_{amb}$ above which the ambipolar diffusion can be safely neglected. Figure \ref{fig:rcc} clearly shows that the most relevant cosmological scales for structure formation are in the region where ideal MHD can be used to treat plasma coupled to neutral baryon matter.

\item Despite the fact that dark energy is considered today to form about $70\%$ of the present Universe, there is strong evidence \citepp{PR03} that it is surely subdominant, or \emph{de facto} negligible in any respect, for the Universe evolution at $z\gtrsim 1$ . Therefore it does not concern the cosmological implementation for the dispersion relation \reff{eq:237}. The situation is very different for dark matter, which forms approximately $80\%$ of the total matter in the present Universe. Dark matter interacts with baryonic, neutral, and ionized matter, respectively, only through the gravitational interaction: its inhomogeneities are greater by a factor of about $20$ than the baryonic perturbations, and therefore they constitute the gravitational skeleton of the structure formation \citepp{KT90,M2011}. However, since the baryon to photon ratio is very small (and almost constant in time) in the background Universe (its value is $\sim 6.1\times 10^{-10}$), the baryons remain tightly coupled to photons, feeling the radiation pressure well after the hydrogen recombination; up to $z\sim 100$ \citepp{LCM12,KT90}. As a consequence, the neutral baryons are prevented from falling into the dark matter gravitational potential by the radiation pressure and therefore they are actually decoupled by the dark matter, but strongly coupled to the ionized and photon components of the Universe.

In this situation, we are clearly neglecting here the dynamics of dark matter perturbations, but we expect to account for the presence of this component by including its contribution in the Universe mass density we introduced in the Jeans length scale 
\reff{eq:240}. Furthermore, we observe that, even when the radiation pressure is suppressed by the Universe expansion ($z<100$), the setting down of baryonic matter in the gravitational potential of dark matter requires a finite time, and therefore the decoupling of the two components remains valid roughly up to $z\gtrsim10$.

More specifically, it is worth noting that the decoupling between photons and baryons depends on the considered scale, since only structures larger than the photon mean free path can really be  coupled to the radiation component: at the recombination age, this scale corresponds to a mass of $\sim 10^{11}$ solar masses. From the point of view of the equation of state, the change regarding the baryon fluid pressure can be interpreted (see \cite{Wein72}) as the passage of the polytropic index from $4/3$ (baryon density behaves as the inverse of the volume, \ie like $a^{-3}$, $a$ being the cosmic scale factor, while the radiation pressure goes as $a^{-4}$), to the value $5/3$, typical of a nonrelativistic fluid. However, to get a quantitative estimate of how the pressure decreases after the recombination and an idea of why it becomes negligible only after $z\sim 100$,  a kinetic evaluation of the sound speed velocity is necessary (see formula \cite{LCM12}). Before recombination (for a scale greater than the photon mean free path), we can write the following expression for the sound speed:
\begin{equation}
v_s^{2} = \frac{c^2}{3}\frac{k_B T_b}{m_p c^2 + 1.5 \times 10^9}\;1.5\times 10^9\;,
\label{reaa1}
\end{equation}
where $m_b$ and $T_b$ denote the baryon mass and temperature, respectively ($k_B$ being the Boltzmann constant). As long as the temperature of the baryon, coinciding with the photon one, remains sufficiently large, this value is close to that of an ultra-relativistic fluid, that is, $v_s = c/\sqrt{3}$. After the recombination, the squared sound speed velocity reads
\begin{equation}
v_s^2=\frac{5}{3}\frac{k_BT_b}{m_b}\;,
\label{reaa2}
\end{equation}
which is a typical nonrelativistic value, some order of magnitude smaller than the estimate \eref{reaa1}. The crucial point is that, up to $z\sim 100$, the baryon and photon temperatures remain essentially equal, while for $z<100,$ baryons rapidly cool as $a^{-2}$ (instead as $a^{-1}$). Only after this time are baryons free particles, and they start to fall in the potential well of the dark matter. For $z>100$, the residual pressure associated with \eref{reaa2} is responsible for acoustic oscillations of the baryon density and the Jeans scale selects which perturbations increase or oscillate. For $z<100$, the sound velocity drastically decreases and the density by which the Jeans scale is calculated increases because the dark matter contribution must be included. As a result, the Jeans threshold scale significantly diminishes, while some perturbations enter the nonlinear dynamical regime.

\end{enumerate}

Even considering the well-posed points above, the implementation of the present investigation of the anisotropic MRI-Jeans instability in a cosmological setting remains valid only on a rather qualitative level. Nonetheless, the obtained results encourage more careful analytical and numerical (maybe N-body) studies to clarify whether or not the formation of filament-like structures across the Universe can be explained with the intrinsic anisotropy of differentially rotating and small magnetized primordial sites.

In this respect, it is worth noting that the presence of the magnetic field introduces a privileged direction in space, which defines an intrinsic anisotropy of the perturbation dynamics, \ie the angle between the background magnetic field and the perturbation wavenumber enters the dispersion relation. We have already outlined such a property in the case of a homogeneous (cosmological) background; see \cite{LCM12} and \cite{PD12}. Clearly, the direction of propagation selects different contributions due to perturbed magnetic pressure and tension, respectively. In particular, the perturbed magnetic pressure depends on the angle mentioned above (it depends directly on the angle between $\boldsymbol{B}_0$ and $\boldsymbol{B}_1$, but the latter is orthogonal to the wavenumber because it has vanishing divergence) and this makes the response due to the magnetic field contrasting the gravitational force intrinsically anisotropic: perturbations propagating along the background magnetic field only provide tension, and, in that direction, the thermostatic pressure alone prevents gravitational collapse (we can speak of pure acoustic oscillations when stable modes are concerned). However, perturbed magnetic pressure has an intrinsic anisotropy and a positive or negative sign: it can support and contrast the ordinary pressure, altering the value of the Jeans scale considered here (for stable modes, we can speak of fast and slow magneto-acoustic oscillations, respectively).

Nonetheless, it is clear that, in the parameter region where MRI is suppressed in favor of the Jeans mode, the anisotropy due to the magnetic field essentially vanishes and the resulting anisotropic growth of the perturbations is due to the intrinsic anisotropy of the background profile, \ie a privileged direction exists because a differential rotation of the system is taken along a given axis. In this respect, our background is not really a homogeneous one and we suggest that it must be cosmologically interpreted as a rotating primordial overdensity, on which we are studying the character of the Jeans instability. As already mentioned, the nonsteady behavior of the cosmological background is a minor feature here, because the considered overdensity can be close to the nonlinear regime, almost disconnected from the Universe expansion, which is removed by the internal bound energy, or negligible on a sufficiently small spatial sub-scale.

We conclude by observing that the aim of this section is only to set up the conceptual framework of a more rigorous and expectedly numerical analysis based on a real cosmological background, and discussing in some detail the linear dynamics of the coupled plasma and baryonic fluids. Since the magnetic field takes a very small value, as dictated by the cosmic microwave background radiation constraint $B_0\lesssim10^{-9}$G, and its energy decays like a radiation component, we could, in first approximation, reliably neglect its influence on the background metric of the expanding Universe (however, for an overdense region, the values taken by $B_0$ can be significantly larger; see \cite{VTP05}). Therefore, an interesting upgrade of our study in combination with a realistic cosmological setting could be ensured by the dynamics of a two-fluid system (plasma and neutral baryons), moving on a Tolmann-Bondi nonstationary and radially symmetric background \citepp{M2011}. The presence of a magnetic field and  ambipolar coupling would make this problem a subtle numerical study (maybe analytical under suitable simplifications) of general relativistic MHD. Nonetheless, we are confident that the qualitative scenario inferred here, together with the order of magnitude of our estimates, will survive in such a refinement of the problem. Clearly, when implementing the present scenario within an appropriate cosmological setting, it would be relevant to provide a precise characterization of the role played, after recombination, by the radiation pressure profile in the stability of the different cosmological scales.

\section{Conclusions}
We have analyzed the local stability of a self-gravitating and differentially rotating magnetized cloud in order to outline how the MHD effects influence the value of the Jeans length in this structure. In other words, we consider an axisymmetric and weakly magnetized system, in which the vertical shear is assumed to be small enough to neglect its stratification. This scheme is the basic paradigm with which the role played by MRI on the growth of gravitational instability is studied and it allows to properly estimate the contributions from magnetic pressure and tension in fixing the precise value of the Jeans critical length.

After a schematic characterization of the background system, we have written down the linear perturbation dynamics and determined the dispersion relation associated to the local linear evolution of the system. We have analyzed specific regimes and then we have outlined the dependence of the dispersion relation on the direction of the perturbation propagation with respect to that of the magnetic field background. We have subsequently studied such anisotropy properties of the dispersion relation concerning the evolution of a spherical overdense region. We have shown that the shape of the perturbation is deeply deformed during its evolution and that it becomes intrinsically anisotropic (up to order of unity), although the anisotropy of the density contrast cannot exceed a few percent, according to the considered linear regime. However, we have clarified how such anisotropy is essentially an intrinsic effect of the Jeans instability since the largest spatial scales, at which it appears to dominate, grow faster in time with respect to the smaller scales where MRI is relevant.

The idea of the Jeans mechanism for the structure fragmentation consists in the decreasing value of the Jeans length as the collapse proceeds and the structure average density increases. The obtained results suggest that such a mechanism is influenced by differential rotation and by the magnetic field too: as the fragmentation of the original background structure evolves, we can infer that the anisotropy degree increasingly affects the substructure shape, generating a filament-like class of subsystems.\vfill\eject

The scenario traced above cannot be used to make generalizations regarding structure formation across the primordial Universe. Nonetheless the common presence of both differential rotation and weak magnetic fields in accreting cosmological structures leads us to infer that the generation of filaments from spherical unstable profiles could concern specific cosmological sites or even subclasses of structures (especially if connected in a nonlinear phase with hot dark matter gravitational skeletons) favoring anisotropic collapses.



\end{document}